\documentclass[aps,twocolumn,superscriptaddress,showpacs]{revtex4}
\usepackage{times,epsfig,amssymb,amsfonts,amsmath}
\usepackage{bm}
\newcommand{\ket}[1]{|#1\rangle}

\begin{document}

\title{$U$ independent eigenstates of Hubbard model}
\author{Ming-Yong Ye}
\email{myye@fjnu.edu.cn}
\affiliation{Fujian Provincial Key Laboratory of Quantum Manipulation and New Energy Materials, College of Physics and Energy, Fujian Normal University, Fuzhou 350117, China}
\affiliation{Key Laboratory of Quantum Information, University of Science and Technology of China, Chinese Academy of Sciences, Hefei 230026, China}
\affiliation{Fujian Provincial Collaborative Innovation Center for Optoelectronic Semiconductors and Efficient Devices, Xiamen, 361005, China}
\author{Xiu-Min Lin }
\affiliation{Fujian Provincial Key Laboratory of Quantum Manipulation and New Energy Materials, College of Physics and Energy, Fujian Normal University, Fuzhou 350117, China}
\affiliation{Fujian Provincial Collaborative Innovation Center for Optoelectronic Semiconductors and Efficient Devices, Xiamen, 361005, China}
\begin{abstract}
Two-dimensional Hubbard model is very important in condensed matter physics. However it has not been resolved though it has been proposed for more than 50 years. 
We give several methods to construct eigenstates of the model that are independent of the on-site interaction strength $U$.
\end{abstract}
\maketitle

Hubbard model is proposed in 1963 \cite{hubbard}, and is now believed to have some relationship with
 the theoretical mechanism of high temperature superconductivity. In the Hubbard model, except for
 single electron hoping terms, there are also  on-site interactions for electrons occupying the same lattice site.
 The Hubbard model looks simple, which actually is difficult to solve except in one dimension. In 1990 Yang and Zhang
 find that the model has a $SO_4$ symmetry and construct many eigenstates that are independent of the on-site interaction strength $U$ \cite{yangzhang}.
 They believe that they have found all the $U$ independent eigenstates of the model,
 but they do not know how to prove the statement. Here we show that their statement is incorrect by giving many new $U$ independent eigenstates of the Hubbard model.
 Our results are demonstrated in two dimensional Hubbard model, but they can be easily extended to higher dimensions.

The Hubbard model on a periodic two dimensional $L\times L$ lattice where $L$ is even is defined by the Hamiltonian
\begin{equation}
H=-t\sum_{\langle\mathbf{r},\mathbf{r}'\rangle}(a_\mathbf{r}^\dag a_{\mathbf{r}'}+b_\mathbf{r}^\dag b_{\mathbf{r}'})+U\sum_\mathbf{r}a_\mathbf{r}^\dag a_\mathbf{r}b_\mathbf{r}^\dag b_\mathbf{r},
\end{equation}
where $\langle\mathbf{r},\mathbf{r}'\rangle$ means the sum is over the nearest neighbor,
$a_\mathbf{r}$ and $b_\mathbf{r}$ are annihilation operators for spin-up and spin-down electrons in coordinate space, respectively,
and $\mathbf{r}=(x,y)$ designates the  $L\times L$ lattice sites with $x$ and $y$ being integers $0,1,\ldots,L-1$.
The operators obey the usual commutation relations for Fermi operators. The parameter $U$ is the on-site interaction strength,
which is usually positive to represent the Coulomb repulsion between electrons on the same site.
Define the annihilation operators $a_\mathbf{k}$ and $b_\mathbf{k}$ in momentum space as
\begin{equation}
a_\mathbf{k}=\frac{1}{L}\sum_\mathbf{r}a_\mathbf{r}e^{-i\mathbf{k}\cdot \mathbf{r}}, b_\mathbf{k}=\frac{1}{L}\sum_\mathbf{r}b_\mathbf{r}e^{-i\mathbf{k}\cdot \mathbf{r}},
\end{equation}
where $\mathbf{k}=(k_x,k_y)$, $k_x$ and $k_y$ are integers times $2\pi/L$ and are restricted to the range $-\pi$ to $\pi$.
The Hamiltonian $H$ can be rewritten as
\begin{equation}
H=\sum_\mathbf{k}E(\mathbf{k})(a_\mathbf{k}^\dag a_\mathbf{k}+b_\mathbf{k}^\dag b_\mathbf{k})+U\sum_\mathbf{r}a_\mathbf{r}^\dag a_\mathbf{r}b_\mathbf{r}^\dag b_\mathbf{r},
\end{equation}
where $E(\mathbf{k})=-2t(\cos k_x+\cos k_y)$.
It is difficult to construct eigenstates of $H$ when neither $t$ nor $U$ is zero.
An important progress is made in 1989, Yang construct many eigenstates of $H$ through a mechanism called $\eta$ pairing \cite{yang}.
Then in the following year, based on the $\eta$ pairing,
Yang and Zhang show that $H$ has a $SO_4$ symmetry and construct more eigenstates of $H$ \cite{yangzhang}. To our knowledge, from then on no new eigenstates of $H$ are constructed.

It is necessary to give an introduction to the eigenstates constructed by Yang and Zhang \cite{yangzhang}, before presenting our new eigenstates.
Their result is
described through spin operators and pseudospin operators. The spin operators are
 \begin{equation}
S_+=\sum_\mathbf{r}a_\mathbf{r}^\dag b_\mathbf{r},S_-=S_+^\dag,S_z=\frac{1}{2}\sum_\mathbf{r}(a_\mathbf{r}^\dag a_\mathbf{r}-b_\mathbf{r}^\dag b_\mathbf{r}).
\end{equation}
The pseudospin operators are
\begin{equation}
J_+=\sum_\mathbf{r}e^{i \bm{\pi} \cdot \mathbf{r}}a_\mathbf{r}^\dag b_\mathbf{r}^\dag,J_-=J_+^\dag,J_z=\frac{1}{2}(N-M),
\end{equation}
where $\bm{\pi}=(\pi,\pi)$, $N=\sum_\mathbf{r}(a_\mathbf{r}^\dag a_\mathbf{r}+b_\mathbf{r}^\dag b_\mathbf{r})$, $M=L^2$ is the total number of the lattice sites, and
$J_+$ is the $\eta$ pairing operator constructed by Yang \cite{yang}. The operators
$H$, $S^2$, $S_z$, $J^2$ and $J_z$ commute with each other and have common eigenstates, where
\begin{equation}
S^2=S_+S_-+S_z^2-S_z, J^2=J_+J_- +J_z^2-J_z.
\end{equation}
Suppose $\ket{\Theta}$ is a common eigenstate of the operators $H$, $S^2$, $S_z$, $J^2$ and $J_z$ with eigenvalues $E$, $s(s+1)$, $s_z$, $j(j+1)$ and $j_z$, respectively, then $J_+^mS_+^n\ket{\Theta}$
is also a common eigenstate with eigenvalues $E+mU$, $s(s+1)$, $s_z+n$, $j(j+1)$ and $j_z+m$, respectively, for $m\leq j-j_z$ and $n\leq s-s_z$.
Yang and Zhang consider the state
\begin{equation} \label{xxx}
\ket{\Upsilon}=b_{\mathbf{k}_1}^\dag b_{\mathbf{k}_2}^\dag\ldots b_{\mathbf{k}_{N_b}}^\dag \ket{0}, N_b=0, 1, \ldots, M,
\end{equation}
where $\ket{0}$ is the vacuum state, which is obviously  a common eigenstate of the operators $H$, $S^2$, $S_z$, $J^2$ and $J_z$. Note that $S_-\ket{\Upsilon}=0$ and $J_-\ket{\Upsilon}=0$, for $\ket{\Upsilon}$ there is
\begin{equation}
s=-s_z=\frac{1}{2}N_b, j=-j_z=\frac{1}{2}(M-N_b).
\end{equation}
Therefore, for a given $N_b$, the states $J_+^mS_+^n\ket{\Upsilon}$ are common eigenstates of $H$, $S^2$, $S_z$, $J^2$ and $J_z$ for $m=0,1,\ldots,M-N_b$ and $n=0,1,\ldots,N_b$.
These are the eigenstates constructed by Yang and Zhang, which are obviously $U$ independent and are believed to be the only $U$ independent eigenstates of $H$ \cite{yangzhang}.
From the way they construct eigenstates, we know that if an eigenstate of $H$ with no double occupation in coordinate space is not an eigenstate of $S^2$, it will
be a new eigenstate that is not found by them, because their eigenstates with no double occupation, i.e., the states $S_+^n\ket{\Upsilon}$ for different $n$ and $\ket{\Upsilon}$, and their superpositions,  are always eigenstates of $S^2$ when the number of electrons is fixed.
\begin{figure}[ht]
\centering
\includegraphics[width=5cm]{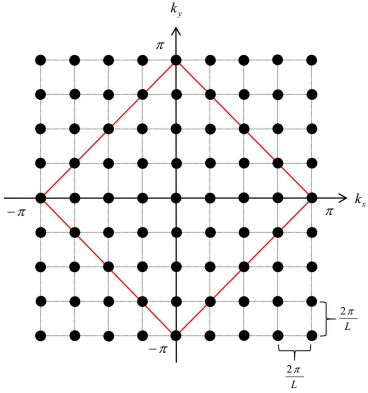}
\caption{Lattice sites in momentum space. The points $\mathbf{k}$ on the red lines have $E(\mathbf{k})=0$, where $L=8$ is shown as an example.}
\end{figure}

Our construction of new eigenstates of $H$ begins with an observation that there are
 many single electron states in momentum space having zero energy.
Figure 1 shows the lattice sites in momentum space, where  the points $\mathbf{k}=(k_x,k_y)$ with $E(\mathbf{k})=0$ are located on the red lines.
It can be found that the number of points $\mathbf{k}$ with $E(\mathbf{k})=0$ grows linearly with $L$.
Note that if $E(\mathbf{k})=0$ there is $E(\mathbf{k}+\bm{\pi})=0$.
Define
\begin{equation} \label{def}
A_{\mathbf{k}\pm}=a_{\mathbf{k}}\pm a_{\mathbf{k}+\bm{\pi}}, B_{\mathbf{k}\pm}=b_{\mathbf{k}}\pm b_{\mathbf{k}+\bm{\pi}},
\end{equation}
and write them using operators in coordinate space
\begin{align}
A_{\mathbf{k}\pm}&=\frac{1}{L}\sum_\mathbf{r}a_\mathbf{r}e^{-i\mathbf{k}\cdot \mathbf{r}}(1 \pm e^{-i\bm{\pi} \cdot \mathbf{r}}), \\
B_{\mathbf{k}\pm}&=\frac{1}{L}\sum_\mathbf{r}b_\mathbf{r}e^{-i\mathbf{k}\cdot \mathbf{r}}(1 \pm e^{-i\bm{\pi} \cdot \mathbf{r}}),
\end{align}
where $e^{-i\bm{\pi} \cdot \mathbf{r}}=e^{-i\pi(x+y)}$, which is $1$ when $x+y$ is even, and $-1$ when $x+y$ is odd.
Therefore
$A_{\mathbf{k}+}$ ($B_{\mathbf{k}+}$) is only a linear combination of $a_\mathbf{r}$ ($b_\mathbf{r}$) with $\mathbf{r}=(x,y)$ indicated by black points
in Fig. 2,
while  $A_{\mathbf{k}-}$ ($B_{\mathbf{k}-}$) is only a linear combination of $a_\mathbf{r}$ ($b_\mathbf{r}$) with $\mathbf{r}=(x,y)$ indicated by red points
in Fig. 2. Now we make the statement that if the state
\begin{equation} \label{main}
\ket{\Psi}=A_{\mathbf{k}_1+}^\dag  \ldots A_{\mathbf{k}_{N_a}+}^\dag B_{\mathbf{q}_{1}-}^\dag   \ldots B_{\mathbf{q}_{N_b}-}^\dag \ket{0}
\end{equation}
is not zero, it will be an eigenstate of $H$ with eigenvalue $E=0$, under the conditions $E(\mathbf{k_i})=0$ for $i=1,\ldots,N_a$ and $E(\mathbf{q_j})=0$ for $j=1,\ldots,N_b$.
In the state $\ket{\Psi}$ the $N_a$ spin-up electrons are on the black points and the $N_b$  spin-down electrons are on the red points in Fig. 2, so there is
no double occupation in coordinate space. Therefore the statement is obvious.
The state $\ket{\Psi}$ is $U$ independent and it is  different from the eigenstates given by Yang and Zhang \cite{yangzhang},
because it is not an eigenstate of $S^2$.
 \begin{figure}[ht]
\centering
\includegraphics[width=5cm]{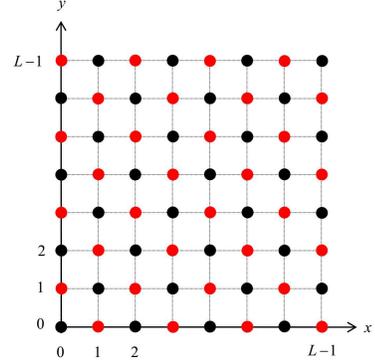}
\caption{Lattice sites in coordinate space. The black points are associated with operators $A_{\mathbf{k}+}$ and $B_{\mathbf{k}+}$, and the red points are associated with operators $A_{\mathbf{k}-}$ and $B_{\mathbf{k}-}$, where  $L=8$ is shown as an example.}
\end{figure}

The above method that we construct eigenstates of $H$ can be slightly modified as follows.
In the definition of $A_{\mathbf{k}\pm}$ and $B_{\mathbf{k}\pm}$ in equation (\ref{def}),
if we change $\bm{\pi}=(\pi,\pi)$ to $\bm{\pi}_1=(\pi,0)$, then
$A_{\mathbf{k}+}$ ($B_{\mathbf{k}+}$) will be only a linear combination of $a_\mathbf{r}$ ($b_\mathbf{r}$) where $x$ is even,
while  $A_{\mathbf{k}-}$ ($B_{\mathbf{k}-}$) will be only a linear combination of $a_\mathbf{r}$ ($b_\mathbf{r}$) where $x$ is odd.
Therefore $\ket{\Psi}$ in equation (\ref{main}) is still a state with no double occupation in coordinate space,
and it will be an eigenstate of $H$ if $E(\mathbf{k_i})=E(\mathbf{k_i}+\bm{\pi}_1)$ for $i=1,\ldots,N_a$ and $E(\mathbf{q_j})=E(\mathbf{q_j}+\bm{\pi}_1)$ for $j=1,\ldots,N_b$,
which is the same as that the $x$ components of the vectors $\mathbf{k_i}$ and $\mathbf{q_j}$ are $\pm \pi/2$.
In a similar way, in the definition of $A_{\mathbf{k}\pm}$ and $B_{\mathbf{k}\pm}$ in equation (\ref{def}),
if we change $\bm{\pi}=(\pi,\pi)$ to $\bm{\pi}_2=(0,\pi)$, the state $\ket{\Psi}$ in equation (\ref{main}) is still an eigenstate of $H$
under the condition that the $y$ components of the vectors $\mathbf{k_i}$ and $\mathbf{q_j}$ are $\pm \pi/2$.

The eigenstates we construct are product states, which   are eigenstates of $H$, $S_z$, $J^2$ and $J_z$, but not eigenstates of $S^2$.
Now we outline a method to obtain common eigenstates of $H$, $S^2$, $S_z$, $J^2$ and $J_z$ from the states we constructed.
We take the state $\ket{\Psi}$ in equation (\ref{main}) as an example and assume $N_a\geq N_b$ without loss of generality.
The state $\ket{\Psi}$ is an eigenstate of $H$, $S_z$, $J^2$ and $J_z$ with $s_z=(N_a-N_b)/2$, $j=(M-N_a-N_b)/2$
and $j_z=-j$,  and it can be written
as a superposition of  eigenstates of $S^2$ with $s=s_z,s_z+1,\ldots,(N_a+N_b)/2$, i.e.,
\begin{equation} \label{y1}
\ket{\Psi}=\sum_{s=s_z}^{(N_a+N_b)/2}\ket{\Psi (s,s_z,j,j_z)},
\end{equation}
where $\ket{\Psi (s,s_z,j,j_z)}$ is a common eigenstate of $H$, $S^2$, $S_z$, $J^2$ and $J_z$.
Note that
\begin{align}
& S_+\ket{\Psi (s,s_z,j,j_z)}=\sqrt{f(s,s_z)}\ket{\Psi (s,s_z+1,j,j_z)}, \\
& S_-\ket{\Psi (s,s_z+1,j,j_z)}=\sqrt{f(s,s_z)}\ket{\Psi (s,s_z,j,j_z)},
\end{align}
with $f(s,s_z)=s(s+1)-s_z(s_z+1)$, there is
\begin{equation}
S_-^n S_+^n\ket{\Psi (s,s_z,j,j_z)}=h(s,s_z,n) \ket{\Psi (s,s_z,j,j_z)}
\end{equation}
with $h(s,s_z,n)=\prod_{i=s_z}^{s_z+n-1}f(s,i)$ when $s_z+n\leq s$, and $h(s,s_z,n)=0$ when $s_z+n > s$.
So we can obtain from equation (\ref{y1}) that
\begin{equation} \label{y2}
S_-^n S_+^n \ket{\Psi}=\sum_{s=s_z+n}^{(N_a+N_b)/2}h(s,s_z,n) \ket{\Psi (s,s_z,j,j_z)}
\end{equation}
for $n=1,\ldots,N_b$,
which together with equation (\ref{y1}) gives us a way to calculate $\ket{\Psi (s,s_z,j,j_z)}$ from $\ket{\Psi}$.
 For any $s\neq (N_a+N_b)/2$, if the calculated $\ket{\Psi (s,s_z,j,j_z)}$ is not zero, it will be an eigenstate of
 $H$ that is not found by Yang and Zhang, because the eigenstates found by them with $N_a+N_b$ electrons and no double occupation
 in  coordinate space have the total spin $s=(N_a+N_b)/2$. When the eigenstate $\ket{\Psi (s,s_z,j,j_z)}$ is obtained,
 more eigenstates of $H$ can be obtained by applying $S_+$, $S_-$, $J_+$ and $J_-$ on it  \cite{yangzhang}.

Now we give an example to demonstrate how we
 calculate the common eigenstates $\ket{\Psi (s,s_z,j,j_z)}$ of $H$, $S^2$, $S_z$, $J^2$ and $J_z$ from the state $\ket{\Psi}$ in equation (\ref{main}). The simplest nontrivial case is $N_a=1$ and $N_b=1$, which leads to $s_z=0$, $j=(M-2)/2$ and $j_z=(2-M)/2$.
From equation (\ref{y1}) and equation (\ref{y2}) we have
\begin{align}
& \ket{\Psi}=\ket{\Psi (0,0,j,j_z)}+\ket{\Psi (1,0,j,j_z)}, \\
& S_-  S_+ \ket{\Psi}=h(1,0,1) \ket{\Psi (1,0,j,j_z)},
\end{align}
where $h(1,0,1)=f(1,0)=2$. Then there is
\begin{equation} \label{y3}
2\ket{\Psi (0,0,j,j_z)}=2\ket{\Psi}-S_-  S_+ \ket{\Psi}.
\end{equation}
Substitute  $\ket{\Psi}=A_{\mathbf{k}_1+}^\dag B_{\mathbf{q}_{1}-}^\dag  \ket{0}$ into equation (\ref{y3}) we get
\begin{equation}
2\ket{\Psi (0,0,j,j_z)}=(A_{\mathbf{k}_1+}^\dag B_{\mathbf{q}_{1}-}^\dag - B_{\mathbf{k}_1+}^\dag A_{\mathbf{q}_{1}-}^\dag) \ket{0},
\end{equation}
which is a singlet state with $s=0$.
Recall the definition of $A_{\mathbf{k}\pm}$ and $B_{\mathbf{k}\pm}$ in equation (\ref{def}),
there is
\begin{equation}
2\ket{\Psi (0,0,j,j_z)}=C^\dag\ket{0}+D^\dag\ket{0},
\end{equation}
where
\begin{align}
& C^\dag=a_{\mathbf{k}_1}^\dag b_{\mathbf{q}_1}^\dag+a_{\mathbf{q}_1}^\dag b_{\mathbf{k}_1}^\dag-a_{\mathbf{k}_1+\bm{\pi}}^\dag b_{\mathbf{q}_1+\bm{\pi}}^\dag-a_{\mathbf{q}_1+\bm{\pi}}^\dag b_{\mathbf{k}_1+\bm{\pi}}^\dag, \\
& D^\dag=a_{\mathbf{k}_1+\bm{\pi}}^\dag b_{\mathbf{q}_1}^\dag+a_{\mathbf{q}_1}^\dag b_{\mathbf{k}_1+\bm{\pi}}^\dag-a_{\mathbf{k}_1}^\dag b_{\mathbf{q}_1+\bm{\pi}}^\dag-a_{\mathbf{q}_1+\bm{\pi}}^\dag b_{\mathbf{k}_1}^\dag.
\end{align}
Both $C^\dag\ket{0}$ and $D^\dag\ket{0}$ are common eigenstates of $H$, $S^2$, $S_z$, $J^2$ and $J_z$ with $s=0$ if they are not zero, due to the conservation of total momentum (mod ($2\bm{\pi}$)). The state $C^\dag\ket{0}$ will be zero only when $\mathbf{k}_1+\bm{\pi}=\mathbf{q}_1$ mod (2$\bm{\pi}$), which is equivalent to
$\mathbf{q}_1+\bm{\pi}=\mathbf{k}_1$ mod (2$\bm{\pi}$). The state $D^\dag\ket{0}$ will be zero only when $\mathbf{k}_1=\mathbf{q}_1$.
Therefore $C^\dag\ket{0}$ and $D^\dag\ket{0}$ can not be both zero.

So far, we construct eigenstates of $H$ only using operators in momentum space with $E(\mathbf{k})=0$, $k_x=\pm \pi/2$ or $k_y=\pm \pi/2$. Now we construct some eigenstates of $H$ using more general operators. We make the statement that the state
\begin{equation} \label{main2}
\ket{\Phi}_{\bm{\pi}}=(a_{\mathbf{k}_1+\bm{\pi}}^\dag b_{\mathbf{q}_1}^\dag -a_{\mathbf{k}_1}^\dag b_{\mathbf{q}_1+\bm{\pi}}^\dag )  \ket{0}
\end{equation}
is an eigenstate of $H$ with eigenvalue $E=0$, under the condition $E(\mathbf{k}_1)=E(\mathbf{q}_1)$. To prove our statement we need to show that $\ket{\Phi}_{\bm{\pi}}$
has no double occupation in coordinate space. This is obvious by writing it using operators in coordinate space
\begin{equation}
\ket{\Phi}_{\mathbf{\pi}}=\frac{1}{L^2}\sum_{\mathbf{r}_1}\sum_{\mathbf{r}_2} a_{\mathbf{r}_1}^\dag b_{\mathbf{r}_2}^\dag e^{i\mathbf{k}_1 \cdot \mathbf{r}_1} e^{i\mathbf{q}_1 \cdot \mathbf{r}_2} g(\mathbf{r}_1,\mathbf{r}_2) \ket{0},
\end{equation}
where $g(\mathbf{r}_1,\mathbf{r}_2)=e^{i\bm{\pi} \cdot \mathbf{r}_1}-e^{i\bm{\pi} \cdot  \mathbf{r}_2}$, which is zero when $\mathbf{r}_1=\mathbf{r}_2$.
When $\mathbf{q}_1\neq \mathbf{k}_1$, the state $\ket{\Phi}_{\bm{\pi}}$ is not an eigenstate of $S^2$, thus it is not an eigenstate of $H$
found by Yang and Zhang.
 When $\mathbf{q}_1=-\mathbf{k}_1$ mod (2$\bm{\pi}$), the condition $E(\mathbf{k}_1)=E(\mathbf{q}_1)$ is satisfied, but the state $\ket{\Phi}_{\bm{\pi}}$
 has already been found by Yang because it has the total momentum $\bm{\pi}$ \cite{yang}. However, for any $\mathbf{k}_1=(k_{1x},k_{1y})\neq \mathbf{0}$  there is $\mathbf{q}_1 \neq \pm\mathbf{k}_1$ mod (2$\bm{\pi}$) and satisfying
  $E(\mathbf{k}_1)=E(\mathbf{q}_1)$, because $\mathbf{q}_1=(q_{1x},q_{1y})$ can be $(k_{1y},k_{1x})$, $(k_{1y},-k_{1x})$, $(-k_{1y},k_{1x})$, $(-k_{1y},-k_{1x})$, $(-k_{1x},k_{1y})$ and $(k_{1x},-k_{1y})$.
Therefore $\ket{\Phi}_{\bm{\pi}}$ represents some new eigenstates of $H$.
We note that in the definition of $\ket{\Phi}_{\bm{\pi}}$ in equation (\ref{main2}),
if we change $\bm{\pi}=(\pi,\pi)$ to $\bm{\pi}_1=(\pi,0)$, then the new state $\ket{\Phi}_{\bm{\pi}_1}$ will be an eigenstate of $H$ under the condition
$k_{1x}=\pm q_{1x}$.
Similarly, in the definition of $\ket{\Phi}_{\bm{\pi}}$ in equation (\ref{main2}),
if we change $\bm{\pi}=(\pi,\pi)$ to $\bm{\pi}_2=(0,\pi)$, then the new state $\ket{\Phi}_{\bm{\pi}_2}$ will be an eigenstate of $H$ under the condition
$k_{1y}=\pm q_{1y}$.

The above method to construct eigenstates of $H$ can be generalized as follows.
Consider the state
\begin{equation} \label{main3}
\ket{\Phi}_{\bm{\alpha}}=(a_{\mathbf{k}_1+\bm{\alpha}}^\dag b_{\mathbf{q}_1}^\dag -a_{\mathbf{k}_1}^\dag b_{\mathbf{q}_1+\bm{\alpha}}^\dag )\ket{0},
\end{equation}
which has no double occupation in coordinate space as $\ket{\Phi}_{\bm{\pi}}$. It will be an eigenstate of $H$ if the condition
\begin{equation}
E(\mathbf{k}_1+\bm{\alpha})+E(\mathbf{q}_1)=E(\mathbf{k}_1)+E(\mathbf{q}_1+\bm{\alpha})
\end{equation}
is satisfied. When $\mathbf{k}_1=\mathbf{q}_1$ the condition is satisfied, but the state has be found by Yang and Zhang because it is an eigenstate of $S^2$ with $s=1$ \cite{yangzhang}. However, when $\mathbf{k}_1\neq \mathbf{q}_1$ the condition can be satisfied in the following cases:
(1) $\bm{\alpha}=(\alpha_x,0)$, $k_{1x}=q_{1x}$, $k_{1y}=-q_{1y}$;
  (2) $\bm{\alpha}=(0,\alpha_y)$, $k_{1x}=-q_{1x}$, $k_{1y}=q_{1y}$;
  (3) $\bm{\alpha}=(\alpha_x,\alpha_y)$ with $\alpha_x=\alpha_y$, $k_{1x}=q_{1y}$, $k_{1y}=q_{1x}$;
  (4) $\bm{\alpha}=(\alpha_x,\alpha_y)$ with $\alpha_x=-\alpha_y$, $k_{1x}=-q_{1y}$, $k_{1y}=-q_{1x}$.
Therefore $\ket{\Phi}_{\bm{\alpha}}$ can represent some new eigenstates of $H$ that are not found before.
Note that in the above four cases, there are $E(\mathbf{k}_1)=E(\mathbf{q}_1)$, $E(\mathbf{k}_1+\bm{\alpha})=E(\mathbf{q}_1+\bm{\alpha})$ and $\mathbf{k}_1+\mathbf{q}_1$ is proportional the vector $\bm{\alpha}$. When $\mathbf{k}_1+\mathbf{q}_1+\bm{\alpha}=\bm{\pi}$ mod ($2 \bm{\pi}$), the state has already been
found by Yang because it has a total momentum $\bm{\pi}$ \cite{yang}, however this can only be occasionally happened in the above case (3) and (4).

Some new eigenstates of $H$ can be constructed in the spirit of $\eta$ pairing.
Suppose $d$ is an integer and define the operators
\begin{equation} \label{main7}
G_{d}=\sum_{k_x} e^{ik_x d}b_{(k_x,\pi-k_x)}=\sum_x e^{-i\pi (x-d)}b_{(x,x-d)},
\end{equation}
where $b_{(k_x,\pi-k_x)}$ is the operator $b_{\mathbf{k}}$ in momentum space with $\mathbf{k}=(k_x,\pi-k_x)$,
and $b_{(x,x-d)}$ is the operator $b_{\mathbf{r}}$ in coordinate space with $\mathbf{r}=(x,x-d)$.
Define
\begin{equation}
F_d (k_x,k_y)=e^{ik_x d} a_{(k_x,k_y)}-e^{ik_y d} a_{(k_y,k_x)},
\end{equation}
where $a_{(k_x,k_y)}$ and $a_{(k_y,k_x)}$ are operators $a_{\mathbf{k}}$ in momentum space with $\mathbf{k}=(k_x,k_y)$
and $\mathbf{k}=(k_y,k_x)$, respectively. $F_d(k_x,k_y)$ can be rewritten as
\begin{equation}
F_d(k_x,k_y)=\frac{1}{L}\sum_{\mathbf{r}}(e^{-i (k_x (x-d)+k_y y)}-e^{-i (k_y (x-d)+k_x y})a_{\mathbf{r}},
\end{equation}
where $a_{\mathbf{r}}$ is an operator in coordinate space with $\mathbf{r}=(x,y)$. Note that $G_{d}$ is
a linear combination of $b_{\mathbf{r}}$ with $y=x-d$, and $F_d(k_x,k_y)$ is
a linear combination of $a_{\mathbf{r}}$ with $y\neq x-d$, therefore if
\begin{equation}
\ket{\Omega}_d=F_d (k_{x_1},k_{y_1})^\dag F_d (k_{x_2},k_{y_2})^\dag \ldots F_d (k_{x_n},k_{y_n})^\dag G_{d}^\dag  \ket{0}
\end{equation}
is not zero, it will be an eigenstate of $H$ because there is no double occupation in coordinate space.  Note that $\ket{\Omega}_d$ is not an eigenstate of $S^2$, therefore it is an eigenstate of $H$ not found by Yang and Zhang \cite{yangzhang}.
We emphasize that the number of electrons in $\ket{\Omega}_d$ can be of the order $L^2$, which is different from the eigenstates we construct in the above.

Before summary we want to give a revisit to the single electron operator $G_d$ in equation (\ref{main7}), where the integer $d$  controls its pattern in coordinate space, and different $d$ represents different pattern.
This observation can be used to construct eigenstates of $H$.
Define operator $Y_d$ as the same as $G_d$ but using spin-up operator $a_{\mathbf{k}}$ instead of spin-down operator $b_{\mathbf{k}}$, i.e.,
\begin{equation}
Y_{d}=\sum_{k_x} e^{ik_x d}a_{(k_x,\pi-k_x)}=\sum_x e^{-i\pi (x-d)}a_{(x,x-d)},
\end{equation}
where $a_{(k_x,\pi-k_x)}$ is the operator $a_{\mathbf{k}}$ in momentum space with $\mathbf{k}=(k_x,\pi-k_x)$,
and $a_{(x,x-d)}$ is the operator $a_{\mathbf{r}}$ in coordinate space with $\mathbf{r}=(x,x-d)$.
It is obvious that the state
\begin{equation}
\ket{\Gamma}=Y_{d_1}^\dag Y_{d_2}^\dag \ldots Y_{d_{N_a}}^\dag G_{d_{N_a+1}}^\dag G_{d_{N_a+2}}^\dag \ldots G_{d_{N_a+N_b}}^\dag \ket{0}
\end{equation}
is an eigenstate of $H$ when all $d_i$ (mod $L$) are different, because it has no double occupation in coordinate space.
However the state $\ket{\Gamma}$ is not a new eigenstate of $H$, but a superposition of states $\ket{\Psi}$ in equation (\ref{main}),
due to the fact that $Y_{d}$ ($G_{d}$) is a superposition of $A_{\pm}$ ($B_{\pm}$) for different $\mathbf{k}=(k_x,\pi-k_x)$ in equation (\ref{def}).

In summary, we have constructed many new $U$ independent eigenstates of $H$ of the Hubbard model in two dimension, and
presented a method to obtain common eigenstates of $H$, $S^2$, $S_z$, $J^2$ and $J_z$ from the states we constructed.
More new eigenstates can be constructed from the states we obtained with the method given by Yang and Zhang according to the $SO_4$ symmetry of the Hubbard model \cite{yangzhang}.
Our results can be easily generalized to higher dimensions.


M. Y. Ye would like to thank Z. J. Yao for helpful discussions about Hubbard model and M. X. Shen for drawing the figures. This work was supported by the National Natural Science Foundation of China (Grant No. 61275215), Fujian Provincial College Funds for Distinguished Young Scientists (Grant No. JA14070) and Natural Science Foundation of Fujian Province (Grant No. 2016J01008, 2016J01009).

\end{document}